# Origin of Relaxor Behavior in Barium Titanate Based Lead-Free Perovskites

*Vignaswaran Veerapandiyan, Maxim N. Popov, Florian Mayer, Jürgen Spitaler, Sarunas Svirskas, Vidmantas Kalendra, Jonas Lins, Giovanna Canu, Maria Teresa Buscaglia, Marek Pasciak, Juras Banys, Pedro B. Groszewicz, Vincenzo Buscaglia, Jiri Hlinka, Marco Deluca[*]*

V. Veerapandiyan, M. N. Popov, F. Mayer, J. Spitaler, M. Deluca
Materials Center Leoben Forschung GmbH, Roseggerstrasse 12, 8700 Leoben, Austria
E-Mail: marco.deluca@mcl.at

S. Svirskas, V. Kalendra, J. Banys
Faculty of Physics, Vilnius University, Sauletekio al. 9, 2040 Vilnius, Lithuania

J. Lins, P. B. Groszewicz
Institute of Physical Chemistry, Technische Universität Darmstadt, 64287, Darmstadt, Germany

P. B. Groszewicz
Department of Radiation Science and Technology, Delft University of Technology, Delft 2629JB, Netherlands.

G. Canu, M. T. Buscaglia, V. Buscaglia
CNR-ICMATE, Institute of Condensed Matter Chemistry and Technologies for Energy, National Research Council of Italy, Via de Marini 6, 16149, Genoa, Italy

M. Pasciak, J. Hlinka
Institute of Physics, Academy of Sciences of the Czech Republic, Na Slovance 2, Praha 8, Czech Republic

**Keywords**
Barium titanate, relaxors, Dielectric spectroscopy, Raman spectroscopy, Nuclear Magnetic Resonance, Density Functional Theory

**Abstract**

It is well known that disordered relaxor ferroelectrics exhibit local polar correlations. The origin of localized fields that disrupt long range polar order for different substitution types, however, is unclear. Currently, it is known that substituents of the same valence as $Ti^{4+}$ at the B-site of barium titanate lattice produce random disruption of Ti-O-Ti chains that induces relaxor behavior. On the other hand, investigating lattice disruption and relaxor behavior resulting from substituents of different valence at the B-site is more complex due to the simultaneous occurrence of charge imbalances and displacements of the substituent cation. The existence of an effective charge mediated mechanism for relaxor behavior appearing at low (< 10%) substituent contents in heterovalent modified barium titanate ceramics is presented in this work. These results will add credits to the current understanding of relaxor behavior in chemically modified ferroelectric materials and also acknowledge the critical role of defects (such as cation vacancies) in lattice disruption, paving the way for chemistry-based materials design in the field of dielectric and energy storage applications.

## 1. Introduction

Recent demands for miniaturized energy storage systems highlight the need for devices combining high power with high recoverable energy density, in order to accommodate rapid energy intake from ambient sources and to provide long-term energy supply [1,2]. Ferroelectric ceramic capacitors have intrinsically high-power density due to fast charge reorientation mechanisms under E-field application. In pure $BaTiO_3$ (BTO) perovskites, this is provided by the long-range correlation of B-site ($Ti^{4+}$) cation displacements. This correlation encompasses also strains up to the microscale, and is embodied by the appearance of ferroelectric domains [3]. Upon field cycling, as needed for charging-discharging the capacitor, ferroelectric domains switch in the direction of the applied field, thereby dissipating the associated elastic energy as heat. These losses are one of the main origins of the low recoverable energy density in this class of materials [4].

One often pursued strategy to increase the recoverable energy density of ceramic capacitors is thus to avoid elasticity-driven losses by disembodying electric charge from elastic strain through the disruption of the long-range correlation of $Ti^{4+}$ displacements [2]. When BTO is substituted at the perovskite B-site, in fact, the correlation of Ti-O-Ti chains is broken [5]; consequently, ferroelectric domains cease to permeate the whole lattice and are rather confined to nanoscale polar regions, the size, correlation and distribution of which depends on substituent type and concentration [6]. The absence of long-range strain correlation limits the losses associated with elastic energy and slims down the polarization-electric field (P-E) hysteresis loop, thereby increasing the recoverable energy density [2], provided that the achievable polarization at the maximum applicable electric field remains closer to (or is higher than) that of pure BTO. There is thus a delicate equilibrium between disruption of ferroelectric long-range order and the retention of a high permittivity level that has to be considered when designing new compositions for high energy density ceramic capacitors. Explaining the nature of the polarization disruption for different chemical modifications of BTO is thus of paramount importance for tuning macroscopic material properties such as energy density.

Barium titanate can be substituted at the B-site either with homovalent (4+, as for instance in $BaZr_xTi_{1-x}O_3$ - BZT) or heterovalent (5+, like in $BaNb_xTi_{1-x}O_3$ - BNbT) substituent ions. By increasing the substituent content, firstly the temperatures of the phase transitions between the ferroelectric phases increase and merge with the decreasing Curie temperature ($T_c$) in a so-called 'triple point' or 'tricritical point' [5]. For higher substituent contents, a diffuse phase transition (DPT) from rhombohedral ferroelectric to cubic paraelectric phase [7,8] is attained (overlapping of sequential phase transitions). For even higher substitutions, relaxor behavior develops: the material presents Vogel-Fulcher-like frequency dispersion of the permittivity maximum ($T_m$), which - unlike $T_c$ - is no longer related to a structural phase transition [9]. Whereas on the macroscopic scale relaxors exhibit a cubic symmetry, which is incompatible with lattice polarization [10], the structure remains polar at a local scale [11]. The level of substitution for both the DPT and transition to a relaxor state also depends on ionic size and on substituent type (hetero- vs. homovalent) [2,7,12].

Relaxors were initially treated with models based on the local compositional fluctuation resulting from different ions spatially distributed in equivalent crystallographic positions [13]. The most widely accepted theory behind the origin of the frequency dispersed dielectric anomaly is the presence of local polarization correlations at the nanoscale (polar nanoregions, PNR), which form from an uncorrelated high-temperature state at the so-called Burns temperature, $T_B$, and then grow with decreasing T until saturation of polar correlations occurs at a freezing temperature, $T_f$, well below ambient conditions [14]. The existence of PNRs in all relaxor systems and their relationship with chemical order/disorder is debated, and there are recent trends to frame evidence behind relaxor behavior with a more fundamental PNR-free argument [15]. Another approach invokes the presence of random electric fields (RF), with different strength and fluctuation dynamics that should originate from chemical substitution [6]. According to Imry and Ma, different ions in the lattice are subjected to different random electric fields as a result of chemical substitution and associated charge-compensating defects, such as vacancies, and the strength of the electric field plays a crucial role in the appearance of relaxor behavior [16]. This model is known for its fundamental approach in addressing the unclear aspects of relaxors, and can be successfully applied to both homo- and heterovalent substituted systems, without necessarily invoking PNRs. One more common theory is based on the striking similarities of highly substituted relaxors to dipolar glasses [17]. The most recent 'slush model' demonstrated a universal relaxor behavior on different material classes and claimed that the high density of domain walls and especially low angle domain walls that result in strong local electric fields are the major contributor to relaxor behavior [18].

Most of the models developed so far target widely studied lead-based relaxor systems, hence they might need adjustments to be applied to lead-free relaxors based on BTO. In BTO, in fact, the polarization is related to the displacement of the $Ti^{4+}$ cation and relaxor behavior is expected to be B-site driven, whereas in the lead-based case the high permittivity comes from displacements of the A-site $Pb^{2+}$ cation [19]. At present, there is no detailed insight on how relaxor behavior emerges in BTO-based systems in response to different types of substitutions.

In this study we demonstrate that the disruption of the long-range spatial correlation of B-site displacements, which is the driving mechanism for the emergence of relaxor behavior in BTO-based perovskites, has a very different origin if homovalent or heterovalent substituents are concerned. We explain here the role of defects and defect-induced disorder in the onset of relaxor behavior in a heterovalent ($Nb^{5+}$-substituted) BTO system (BNbT), and compare it against the well-studied $Zr^{4+}$-substituted BTO-based homovalent relaxor system (BZT). The discussion we provide is supported by local structural methods, such as Raman and Nuclear Magnetic Resonance (NMR) spectroscopy, evidencing the evolution of lattice disorder upon substitution in both systems, and the presence of defect-related phonon modes in the Raman spectra - calculated by the recently developed spherical averaging method [20] - to reveal the effect of substitution type in the local structure. In addition, Density Functional Theory (DFT) simulations of unit cell volume, strain and electrostatic potential landscape showing the effect of defect dipoles in supercells of the substituted BTO lattice are presented. In summary, this work describes the difference between homovalent and heterovalent relaxors from a combined experimental-theoretical viewpoint, and thus unravels the underlying nature of polarization disruption in chemically modified BTO based materials.

## 2. Results and Discussion
### 2.1. Chemically Modified BTO

As a prime example of how relaxor behavior differs in homovalent- and heterovalent-substituted BTO, we consider systems in which the $Ti^{4+}$ (B-site) cation is substituted by either $Zr^{4+}$ or $Nb^{5+}$. BZT is a homovalent-substituted solid solution where the $Zr^{4+}$ cation has a larger ionic radius than $Ti^{4+}$ (0.72 Å vs. 0.605 Å, respectively [21]). BNbT (with formula bruta $BaNb_xTi_{1-5x/4}O_3$, taking into account a B-site charge compensation scheme) presents heterovalent substitution where $Nb^{5+}$ has a similar ionic radius (0.65 Å [21]) to $Ti^{4+}$. These differences (be it the oxidation states and/or ionic radii) define the predominant mechanism by which these substituents disrupt the BTO lattice continuity: $Zr^{4+}$ will introduce strain, likely without strong contributions from charge distribution, whereas in the case of $Nb^{5+}$, a charge mediated mechanism is likely to play a predominant role. This is also suggested by the necessary presence of charge-compensation schemes for $Nb^{5+}$ incorporation: either Ba or Ti vacancies, represented in Kroger-Vink notation as $V''_{Ba}$ and $V''''_{Ti}$, respectively, must be present in BNbT.

The phase evolution presents significant differences in BZT and BNbT. **Figure 1** shows the phase diagram of both systems highlighting the Curie temperature ($T_c$) for FE compositions and the temperature of relative permittivity maximum ($T_m$) for relaxor compositions, together with dielectric spectroscopy data taken on our samples. The temperature-dependent permittivity at different frequencies is presented to highlight the ferroelectric to relaxor transition, and corresponding results on pure BTO are presented for reference. Our results are in accord with previous findings: Increasing substituent content disrupts ferroelectricity in both systems, as signaled by the shift of the permittivity maximum to lower temperatures and its overall decrease in magnitude. While a diffuse phase transition towards the high-temperature cubic phase is observed for BZT at a substituent content of x=0.10, a lower $Nb^{5+}$ content (x=0.025) is sufficient in case of BNbT. A relaxor behavior, marked by the frequency dispersion of permittivity maximum, is attained for higher substituent contents. Interestingly, the threshold for relaxor behavior is very different in the heterovalent system compared to the homovalent case, with a substitution concentration of x = 0.07 for BNbT and x = ~0.25 in BZT, the latter threshold being observed at similar substituent contents also in other homovalent systems like $BaCe_xTi_{1-x}O_3$ and $BaTi_{1-x}Hf_xO_3$ systems [5,22]. This "early" onset of relaxor behavior is a common trait of heterovalent systems [2], even if substitution is on the A-site like the case of $La^{3+}$ (Ref. [23]).

Since the driving force for relaxor behavior is the breaking of the long-range spatial correlation of $Ti^{4+}$ (B-site) cation displacements in the BTO lattice, we can envisage that different substituents will have a different role in inducing disorder at the perovskite B-site. Recent studies on homovalent-substituted systems, e.g. with $Ce^{4+}$ (Ref. [5]) and $Zr^{4+}$ (Ref. [8]), confirmed that large homovalent cations promote a non-polar (cubic) unit cell, and thus act as Ti-O-Ti chain breaking centers. Since the off-center displacement of $Ti^{4+}$ is related to the amount and distribution of the substituting homovalent cations[24], the latter act as ferroelectric domain pinning centers, effectively breaking down larger domains into smaller polar entities. As soon as the substituent content is increased, these domains are expected to shrink in size down to the atomic scale. Previous studies on heavily substituted homovalent relaxor compositions (like $BaZr_{0.50}Ti_{0.50}O_3$ [25]) suggested in fact that PNRs are centered on $Ti^{4+}$ clusters of few unit cells. At lower substitution level, non-polar regions coincident with the substituents are embedded into a polar matrix as long as there is sufficient long-range correlation between polar Ti-O-Ti chains. It is somewhat clear that in BTO-based systems, the influence of the substituting cations on $Ti^{4+}$ displacements and on their ability to spatially correlate is thus the determining factor for the appearance of relaxor behavior. While for the homovalent case an ionic-size driven "steric hindrance" or "random strain field" effect is easily envisaged, a different mechanism for polarization disruption must exist in heterovalent substituted systems like BNbT, where the similar ionic size between $Ti^{4+}$ and $Nb^{5+}$ precludes a strong strain-driven effect from appearing. Such alternative mechanism shall also explain in particular why the effect of heterovalent substituents is decisively more effective than homovalent ones in disrupting the long-range FE order, as evidenced in Figure 1 from the viewpoint of dielectric properties.

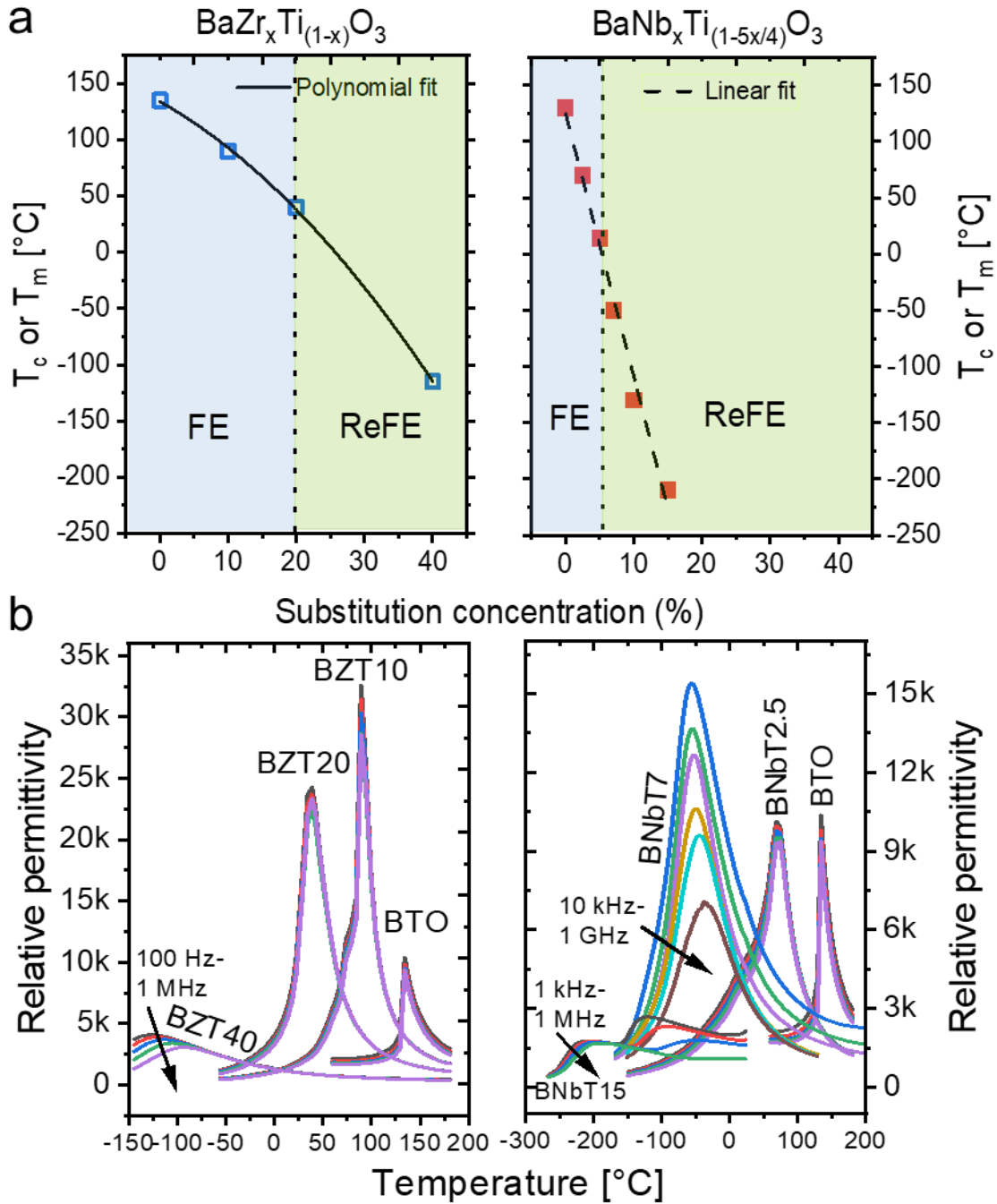

**Figure 1.** Curie temperature (FE) or temperature of permittivity maximum (relaxor) of BZT and BNbT samples, clearly depicting the rapid and early onset of relaxor behavior in BNbT compared to BZT (a). Temperature dependent relative permittivity at different frequencies for BZT and BNbT, demonstrating relaxor behavior by the frequency dispersion of permittivity for 40% Zr and 7% Nb substituent concentration, respectively (b).

## 2.2. Dielectric Relaxation in Heterovalent-Substituted BTO

In heterovalent-substituted BNbT, merging of phase transitions without dielectric dispersion is evident already for BNbT2.5, as shown in Figure 1(b). For increasing substituent concentration (for 7% and above), two dielectric relaxations appear as a function of temperature: one at $T_m$ (relaxor-like) and a second one at higher temperatures, the latter clearly evident only at 15% $Nb^{5+}$ content. **Figure 2** shows a detailed temperature and frequency dependent analysis of dielectric relaxation for the BNbT15 relaxor composition. The second relaxation (marked as region II) occurs in BNbT15 in the temperature range of -150 °C to RT and hence overlaps with the relaxor-like relaxations in BNbT7 and BNbT10, whose $T_m$ are -50 °C and -130 °C respectively, at 1 kHz. For this reason, it is clearly visible only in BNbT15 together with the relaxor-like dispersion of $T_m$ (marked as region I) with dispersion that extends at least up to 30 GHz, as shown in Figure 2(a).

The progressive increase in real part of permittivity with decreasing temperature at lower frequencies indicates another competing relaxation process (cf. Figure 2(b)). As shown in Figure 2(c), for almost all temperatures within the region II, the breadth of the imaginary permittivity peak did not change with decrease in temperature, which is in contrast to a relaxor-like dispersion, in which the relaxation progressively broadens with decreasing temperature [26]. Hence, the relaxation in region II is not related to dynamic polarization fluctuations, but rather to a Debye-type relaxation, which could in principle originate either from charge imbalances located on lattice defects [27] or from interfacial (i.e. Maxwell-Wagner type) free charge carriers [28]. More detailed information about the relaxor-like dispersion in region I and the Debye-like relaxation in region II is given in the Supplementary Information.

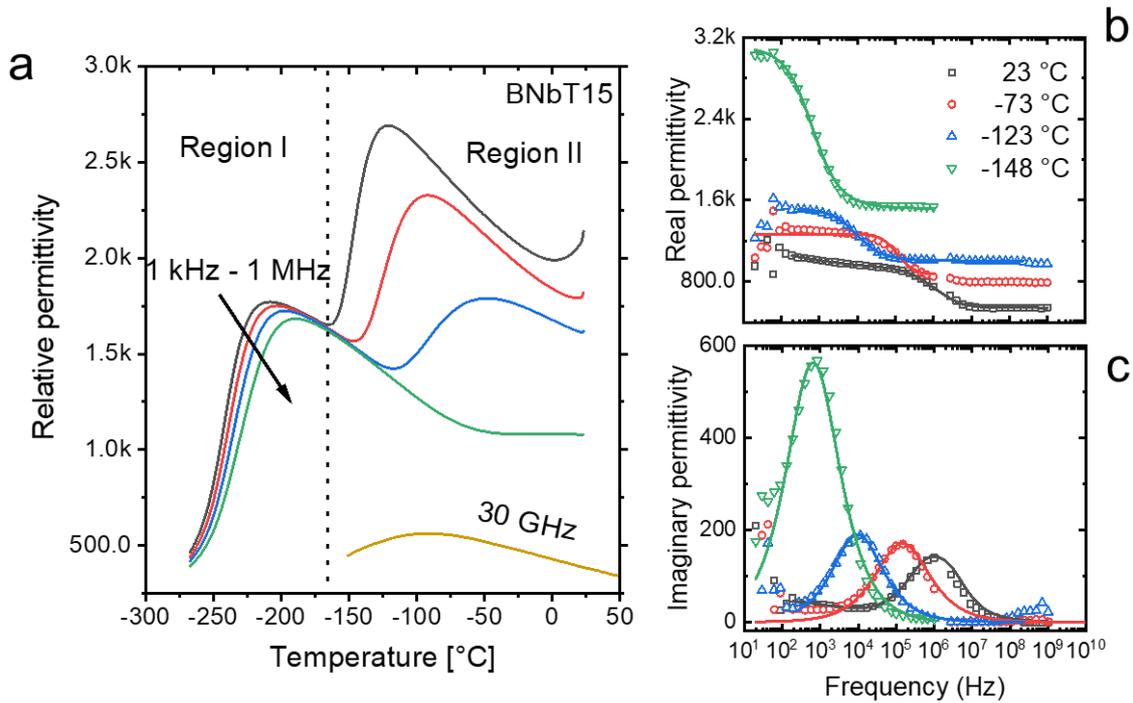

**Figure 2.** (a) Dielectric relaxation of $BaNb_{0.15}Ti_{0.81}O_3$ as a function of temperature at different frequencies highlighting two dielectric relaxations differentiated as region I and region II as marked by a vertical dotted line. The partition line shown in the figure is merely a guidance to distinguish the two relaxations. (b, c) Complex permittivity of $BaNb_{0.15}Ti_{0.81}O_3$ as a function of frequency at different temperatures in region II, highlighting the fact that the breadth of imaginary permittivity peak is unchanged with decreasing temperature, unlike relaxor-like dispersion.

**2.3. Lattice Disorder in Heterovalent-Substituted BTO**

**Figure 3** (a, b) reports the Raman spectra of all investigated compositions of BZT and BNbT. The assignment of Raman modes in pure BTO is described in detail in previous work [20]. From the weakening of the ferroelectric-related mode around 300 cm$^{-1}$ (Ref. [29]), it is evident that in BZT ferroelectric long-range order is lost between 20% and 40% of $Zr^{4+}$ content, whereas in BNbT this occurs at a lower concentration of heterovalent B-site cation, namely between 5% and 7% of $Nb^{5+}$ content, in agreement with dielectric measurements.

In addition to the typical first-order Raman bands of BTO, there are several other bands that can be associated with either $Zr^{4+}$ or $Nb^{5+}$ substitution [8,12,30]. Interestingly, the spectral bands of BNbT are broader than those of BZT (cf. the spectra for 10% substituent content). The spectral width of the Raman bands indicates the correlation length of phonon propagation, and sharp bands with well-defined peak maxima imply a greater correlation length, which if seen in the reciprocal lattice corresponds to pure gamma-point phonons. In contrary, a broad spectral signature suggests phonon activation outside the Brillouin zone center, which in a real lattice is associated with the onset of lattice disorder.

The local structural disorder evidenced by broad Raman bands in BZT and BNbT – stronger in the latter - can be quantified by means of Nuclear Magnetic Resonance (NMR) spectroscopy [31]. In Figure 3 (c, d) we report $^{137}$Ba NMR spectra of BNbT and BZT samples recorded at 133°C and 117°C, respectively. $^{137}$Ba NMR is a sensitive probe to distortions of the local structure, as the reported line width is directly related to gradients of the electric field (EFG), a local structure parameter at the nuclear site. The local structure distortion responsible for an EFG ≠ 0 can be either caused by nearby defects (i.e. substituents) or by symmetry breaking in a specific polymorph or polar structure. To suppress the influence of the latter, and hence quantify the lattice disorder in an unbiased manner, samples were analyzed above $T_c$, as no polar displacements should be observed in the cubic, paraelectric phase. This fact can be illustrated in BTO, for which the linewidth follows the temperature-dependent macroscopic polarization and exhibits a negligible line width above $T_c$ [32], also shown in Figure S4. In short, in the paraelectric state - where electrical polarization (P) do not prevail - BTO displays a narrow linewidth of 600 Hz (FWHM), as shown in Figure 3 (c, d), in which dipolar interactions are the sole contributors to the linewidth with an EFG = 0 (i.e. $C_Q$ = 0).

For BNbT, the $^{137}$Ba NMR linewidth increases from 1.5 kHz in BNbT2.5 to 45.0 kHz in BNbT15, as consequence of increasing structural distortion for higher substituent concentration (Figure 3(d) and Figure S3). The wider lines observed for BNbT in the paraelectric state indicate that, locally, a significant deviation from cubic symmetry is present. An analogous trend is observed for the BZT samples (Figure 3(c)), albeit at a higher substituent content compared to BNbT. Here, it is worth mentioning that $^{137}$Ba NMR spectra of chemically modified samples exhibit a broad asymmetric peak unlike the line shape observed in the ferroelectric phases of BTO [33]. This featureless line shape results from the substitution-induced chemical disorder and the accompanying distribution of EFG components for the local environment of $Ba^{2+}$ across the lattice, which was previously reported for A-site substituted BTO [34].

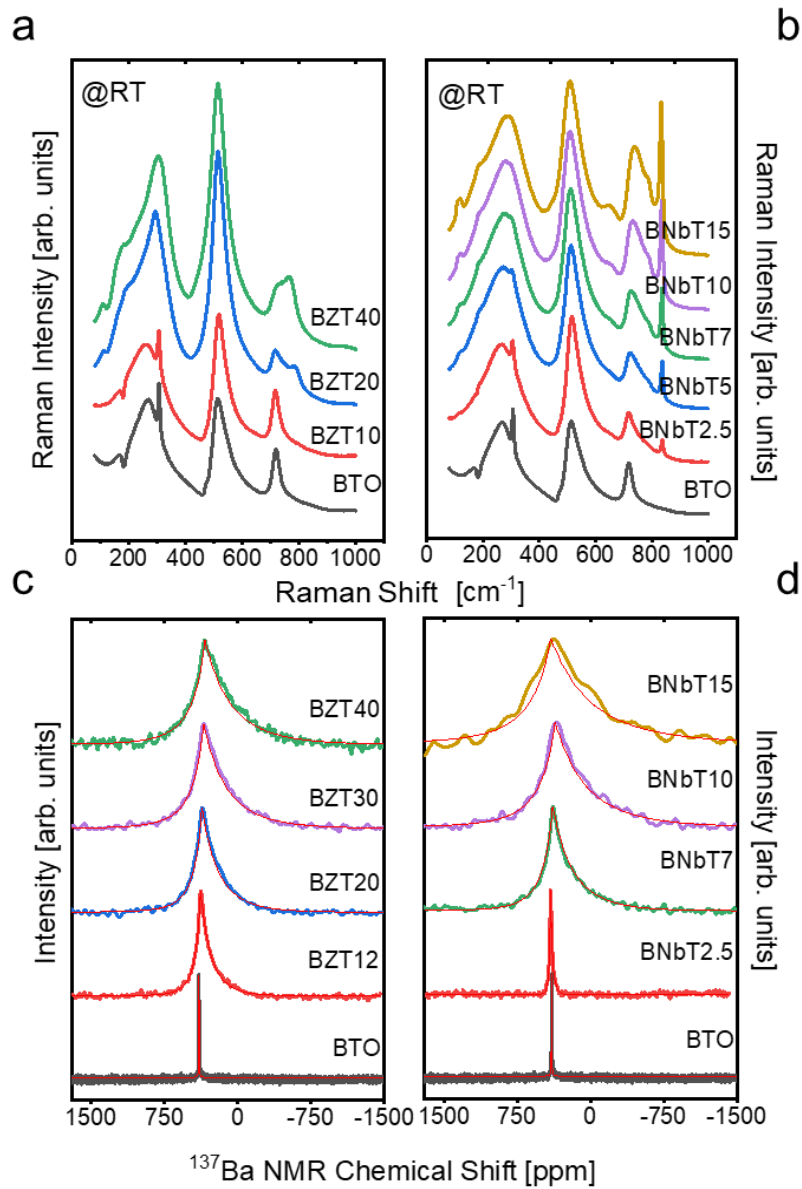

**Figure 3.** Composition dependent Raman spectra of $BaZr_{1-x}Ti_xO_3$ (a) and $BaNb_xTi_{1-5x/4}O_3$ (b) at room temperature. The Raman spectra of BTO at the bottom is added for reference and several substitution-induced Raman bands are evident in both material systems when compared to BTO. The disappearance of the Raman band at ~300 cm$^{-1}$ implies a loss of long-range ferroelectric order, that is evident for >20% of $Zr^{4+}$ and >7% $Nb^{5+}$ concentration. $^{137}$Ba NMR static spectra of $BaZr_{1-x}Ti_xO_3$ (c) and $BaNb_xTi_{1-5x/4}O_3$ (d)

samples recorded at 117 °C and 133 °C, respectively, both above the Curie temperatures of all samples. The presence of broad NMR peaks is related to the existence of disorder in the lattice.

Non-zero EFG above $T_c$ can be directly related to contributions from static and dynamic disorder in the lattice. The Czjzek model expresses the distribution of EFG components in terms of the mean quadrupolar coupling constant ($C_Q$)[35], a parameter that reflects the extent of local structural disorder for the $Ba^{2+}$ site in BZT and BNbT. Considering the fact that spectra are recorded in the paraelectric phase, the mean $C_Q$ value can be interpreted as a measure of the local structural disorder caused by the substituents only, an assumption that can be reinforced by the absence of temperature dependent evolution of linewidth for BNbT samples above $T_c$, as reported in Supplementary Information (cf. Figure S3). **Figure 4** reports the mean $C_Q$ for all investigated compositions, and highlights that $Nb^{5+}$ is much more efficient in disrupting lattice order than $Zr^{4+}$ when substituted at the B-site. While a steep increase in mean $C_Q$ values is observed for small $Nb^{5+}$ contents, a much higher amount of $Zr^{4+}$ is required to produce the same effect. For instance, a 7% $Nb^{5+}$ substitution introduces as much disorder as that of 20% $Zr^{4+}$ in the BTO lattice, as reflected by the mean $C_Q$ of 3917 kHz for BNbT7 which is close to a value of 3537 kHz for BZT20. It is interesting to note that $C_Q$ values above ~4000 kHz mark a threshold for occurrence of relaxor behavior, regardless of substituent type. This value lies slightly above the $C_Q$=2800 kHz observed for room-temperature BTO in the P4mm polymorph [33]. Moreover, these results show that the amount of local structure disorder induced by $Nb^{5+}$ substitution is larger than that induced by $Zr^{4+}$ addition, and confirm the initial supposition on the basis of the Raman linewidths. This striking difference in ability to induce lattice disorder between homovalent and heterovalent cations hints at a completely different mechanism at play in each substitution type.

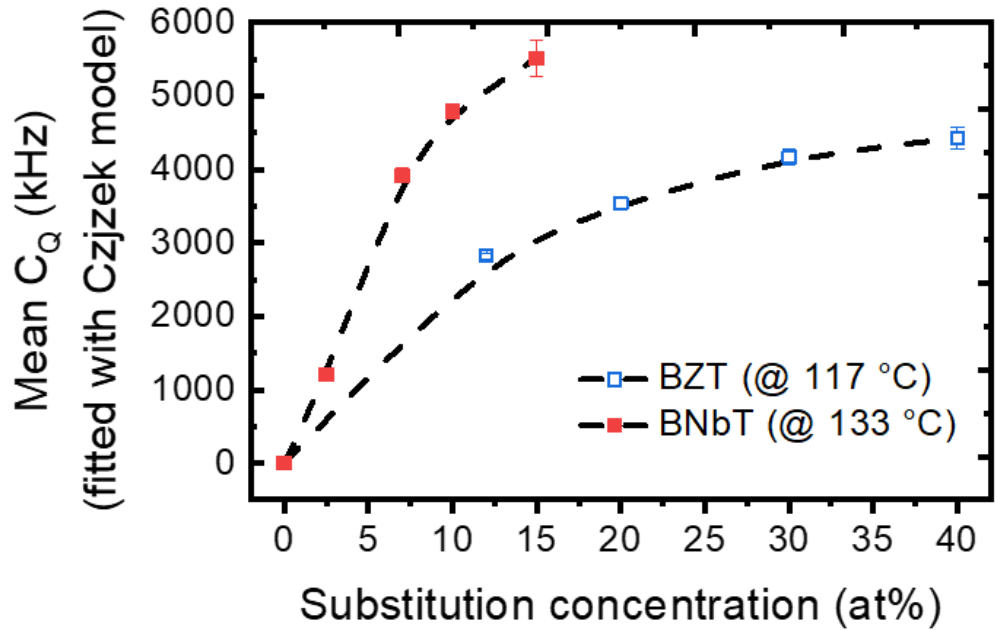

**Figure 4.** The mean $C_Q$ extracted from the $^{137}$Ba NMR for BZT and BNbT samples, depicting a drastic increase in the structural disorder for BNbT (closed symbols) compared to BZT (closed symbols). The mean $C_Q$ values were extracted from spectra recorded at 117 °C and 133 °C for BZT and BNbT samples, respectively, which is above the Curie temperatures of all samples.

**2.4. Charge-Compensating Defects in Heterovalent Relaxors**

From the NMR analysis and the broader Raman modes we evinced that heterovalent substitution induces stronger lattice disorder in BTO compared to the case of homovalent substitution. Furthermore, dielectric spectroscopy data also evidenced the presence of a peculiar dielectric relaxation at temperatures higher than the relaxor-like dispersion, which is ascribed to a Debye-like relaxation compatible with the presence of charge imbalances caused by point defects in the lattice. In this section, we employ a combination of DFT-simulated Raman spectra and DFT calculations of different point defect models to elucidate the nature of these defects. Furthermore, this approach allows us to explore their role in the macroscopic dielectric properties of heterovalent modified BTO by analyzing their effect on the local structure with respect to lattice strain and electric potential and, ultimately, to the disruption of Ti-O-Ti correlations related to electrical polarization.

In **Figure 5**, the Raman spectra of BZT40 and BNbT15, both measured at -185°C, are presented, with some Raman modes highlighted that are associated with lattice defects in chemically substituted BTO systems [36]. In particular, mode 4 (at ~750 cm$^{-1}$) was related previously to the difference in ionic radius of the substituting cation compared to the Ti$^{4+}$ host ion [37]. Following this reasoning, mode 4 should be closer to mode 3 in BNbT than in BZT, which clearly is not the case.

In addition, we notice in BNbT the presence of two extra modes (2 and 5, at ~630 cm$^{-1}$ and ~830 cm$^{-1}$, respectively), one of which (mode 5) was previously reported in both BNbT [12] and La-substituted BaTiO$_3$ [37]. In the BNbT case, mode 5 was associated to a localized BO$_6$ oxygen breathing vibration in the vicinity of Nb$^{5+}$ cations [12]. In the La$^{3+}$ case, on the other hand, this mode was associated with $V_{Ti}''''$ providing charge compensation for the La$^{3+}$ addition [37]. We also recently demonstrated that this mode is absent in dipole-compensated systems (such as BaGa$_x$Ta$_x$Ti$_{1-2x}$O$_3$ and BaGa$_x$Nb$_x$Ti$_{1-2x}$O$_3$), where formation of B-site cation vacancies is limited [36,38]. Mode 5 is also clearly composition-dependent and displays no temperature dependence (cf. Figure 3 & Figure S5), contrary to what would be expected from long-range oxygen sublattice modes. Hence, it is suggested that peak 5 can be attributed to a localized mode originating from charge-compensating B-site defects in heterovalent-substituted BTO. The same suggestion applies to mode 2, which is also present only in heterovalent-substituted BTO. This mode was previously associated to the presence of stress-induced hexagonal BTO phase [29,39], but this may apply only to nanostructured or thin film BTO, and not to the present bulk case.

To clarify the origin of extra modes 2 and 5, we calculated the full Raman spectral signature of both BZT40 and BNbT15 within 3x3x3 supercells, following the spherical averaging method we developed recently based on ab initio Raman-active phonon calculations [20]. In BZT, each supercell was constructed with either Ti$^{4+}$ or Zr$^{4+}$ at the B-site in the right proportion to realize 40% Zr$^{4+}$ amount. Whereas, for BNbT the presence of either 1 $V_{Ti}''''$ and 4 $Nb^{5+}$ at the B-site or 1 $V''_{Ba}$ at the A-site and 2 $Nb^{5+}$ at the B-site was considered for the 3x3x3 supercell. Figure 5 reports the calculated Raman spectra for BZT40 and BNbT15 compared against the respective experimental spectra (at -185°C). Most of the experimental Raman spectra of polycrystalline materials show asymmetric Raman bands due to the convolution of $A_1$/E peaks, LO/TO peaks and oblique phonon dispersion [20] and also significant peak broadening in chemically modified systems because of the lattice disorder, as discussed in the previous section. Hence, the calculated spectra here show spectral signatures that are impossible to be deconvoluted in the measured spectra at -185°C. This notwithstanding, the spectral signature above 600 cm$^{-1}$ - where all the substitution-related Raman modes are present - is clearly replicated in the calculated Raman spectrum. Notably, in BNbT the extra Raman modes 2 and 5 are only seen when supercells were constructed with the $V''''_{Ti}$ charge compensation scheme, which allows to unambiguously attribute their origin to B-site cation vacancies alone. This is in contrast to low donor concentrations and acceptor doping, where results from literature indicate that heterovalent doping is compensated by electrons[40,41]. Following this, we evaluated the sum of the square atomic displacements for the O atoms around the $V''''_{Ti}$ in BNbT15, to highlight their contributions to the Raman-active phonons, as reported in **Figure 6**. From this analysis it is clear that the contribution is stronger in the region of occurrence of the two defect modes (Mode 2 and mode 5 - labelled as "B" and "C" in Figure 6). Actually, the contributions of these oxygens are also pronounced for a mode just below 200 cm$^{-1}$, albeit masked here in experimental spectra by superposition with neighboring peaks. In particular, the sole contributor of mode 5 is the oxygen neighbors of $V''''_{Ti}$ (also evident from the video available in the Supplementary Information) and thus can be unambiguously assigned as a localized $BO_6$ oxygen breathing mode in the vicinity of a $V''''_{Ti}$ (not in the vicinity of $Nb^{5+}$ atoms as previously speculated by Farhi et al. [12]). This confirms the previous reports of this mode in $La^{3+}$ substituted BTO (where $V''''_{Ti}$ are the preferred charge compensation scheme and mode 5 is present) [36,37], and provides an explanation for its absence in dipole-compensated BTO [36,38], where $V_{Ti}''''$ are severely restricted.

On the basis of the evidence for the existence of $V_{Ti}''''$ defects, we note that each vacancy compensates for four $Nb^{5+}$ substituent cations, according to the defect chemical reaction below:

$2Nb_2O_5 \rightarrow 4\ Nb_{Ti}^{*} + V_{Ti}'''' + 10\ O_O^x$

Intuitively, $Nb_{Ti}^{5+}$ shall form clusters around Ti vacancies, due to Coulombic attraction. In an attempt to determine the precise distribution of $Nb^{5+}$ cations in the vicinity of $V_{Ti}''''$, we considered two possible cluster geometries (see Figure S6) and showed that formation energy becomes identical for both considered cluster configurations with increasing supercell size (see Figure S7). This means that the lattice is charge compensated, but locally charge imbalances from lattice defects are present, which may impact on the macroscopic dielectric properties as a function of temperature and applied electric field frequency, irrespective of cluster geometries.

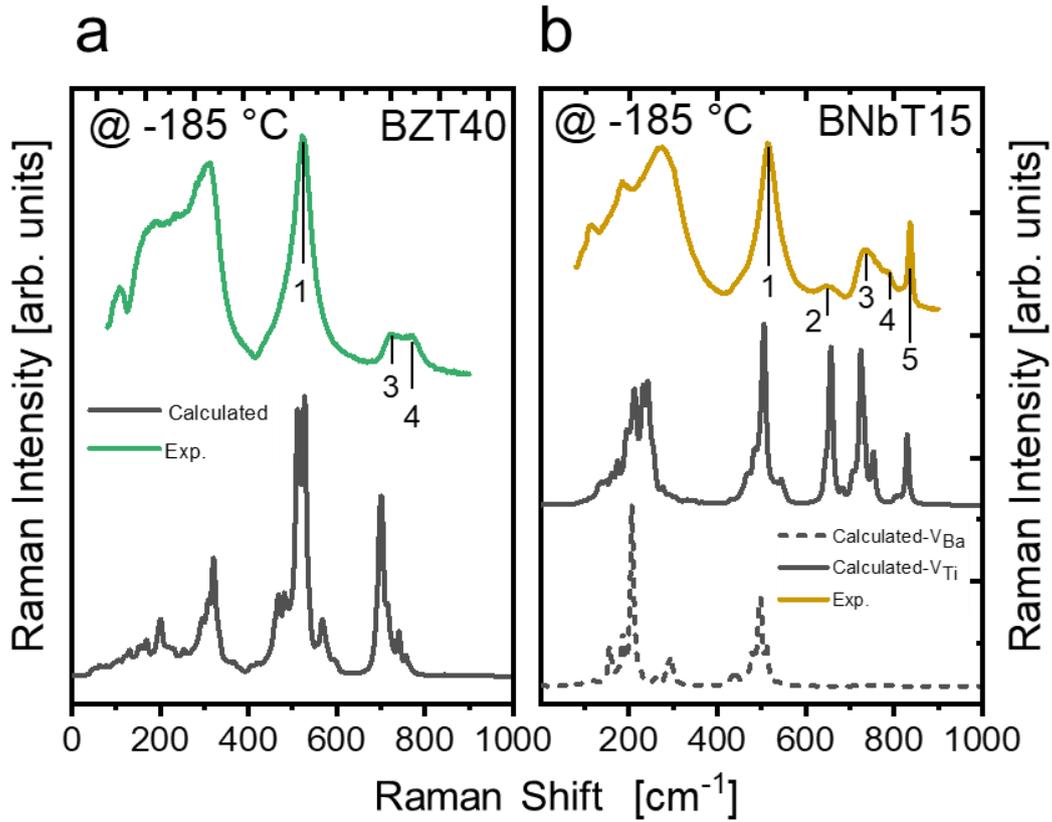

**Figure 5.** Experimental Raman spectra of BZT40 (a) and BNbT15 (b) relaxor systems recorded at -185°C along with the calculated Raman spectra. The experimental data show broad spectral signatures specific to locally disordered relaxor systems. The calculated Raman spectra not only replicate all the experimental Raman bands but are also able to reproduce some of the underlying spectral signatures that are difficult to deconvolute in the experimental data.

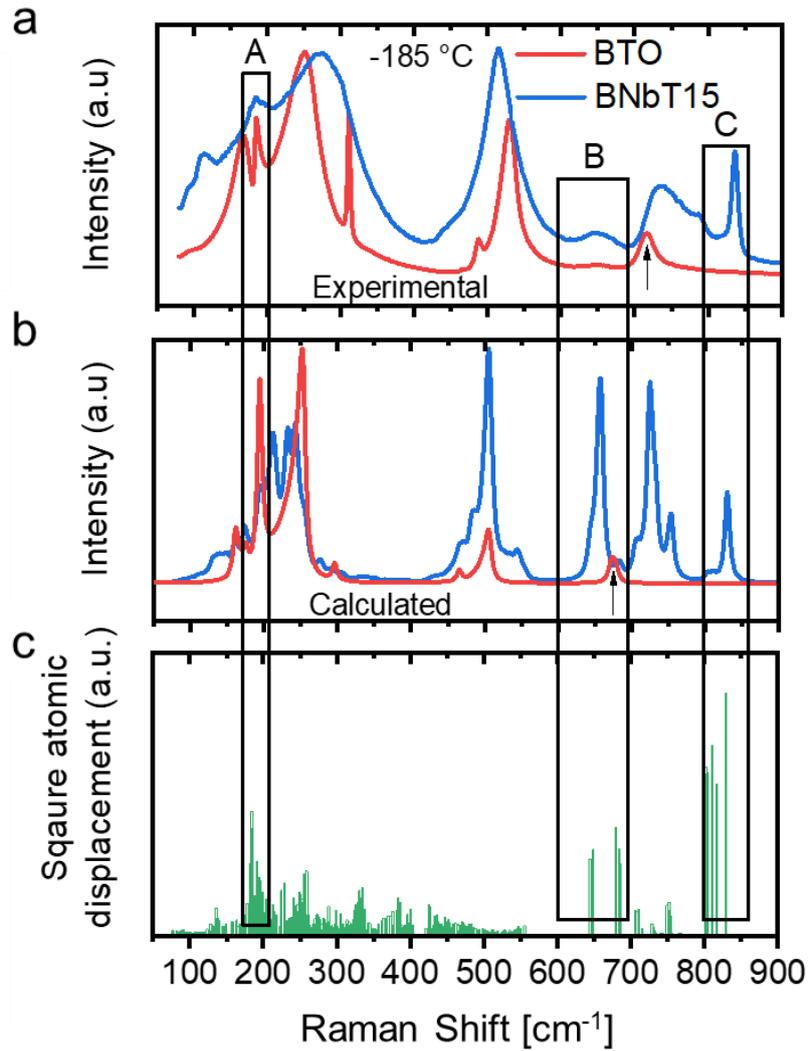

**Figure 6.** Measured Raman intensities for BTO and BNbT15 at -185°C (a) and the calculated Raman spectra of BTO and BNbT (b). (c) Eigenmode analysis for 6 oxygen atoms around the titanium vacancy, where the sum of the square displacement of the 6 oxygen atoms is plotted for each mode. This analysis clearly describes the increased contribution of oxygen atoms to Raman-active phonons near the defect related modes that are marked as B and C.

Further arguments supporting the effect of presence of $V_{Ti}''''$ resulting from charge compensation for BNbT is found by contrasting dielectric relaxation between BZT and BNbT in the highest substituent content studied. Figure 2 and Figure S8 show the temperature dependent permittivity response in relaxor compositions, i.e., BNbT15 and BZT40, respectively. The relaxor behavior is evidenced from the low temperature frequency dispersion ($f$ dependent $T_m$) of $\varepsilon_r$ in both compositions. As evident, there are some discrepancies in the dielectric response between BZT40 (which is in accord with previous literature[7]) and BNbT15, most notably the double relaxation we already discussed. There are however further striking differences in BNbT compared to previous literature [12]. **Figure 7** shows the dielectric permittivity response of BNbT15 at 1 MHz with a Ba vacancy ($V_{Ba}''$-Kroger-Vink notation) as charge compensation scheme (dotted line), extracted from Farhi et al. [12] (henceforth ascribed as BNbT15-$V_{Ba}$), whereas the solid line shows the frequency dependent dielectric permittivity response of BNbT15 from the present work, which involves $V_{Ti}''''$ as charge compensation scheme (BNbT15-$V_{Ti}$). First of all, in BNbT15-$V_{Ba}$, the double relaxation is absent. Then, the $T_m$ at 1 kHz is 63°C lower for BNbT15-$V_{Ti}$ compared to BNbT15-$V_{Ba}$ (cf. Figure 7 (a, b), the $\varepsilon_r$ value is larger for BNbT15-$V_{Ba}$ compared to BNbT15-$V_{Ti}$ and the diffusivity of the $\varepsilon_r$ response is larger for BNbT15-$V_{Ti}$ compared to BNbT15-$V_{Ba}$ (cf. Figure 7 (a). Here it is important to note that although Farhi et al. claim $V_{Ba}''$ as a major charge compensation scheme, it is highly unlikely from a processing standpoint to achieve such high substitution concentration without incorporating $V_{Ti}''''$, and we envisage simultaneous occurrence of both $V_{Ti}''''$ and $V_{Ba}''$ in that case. This can be reaffirmed by the presence of mode 5 in Raman spectra reported in that work [12].

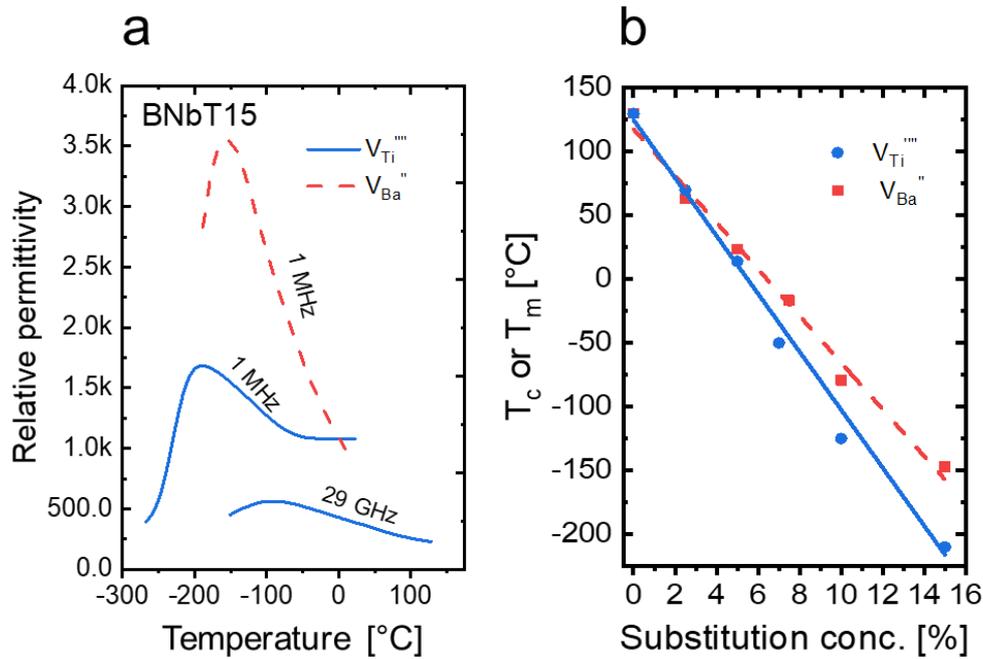

**Figure 7.** Temperature dependent dielectric response of BNbT15 at 1 MHz BNbT15-$V_{Ti}$ (solid line) and BNbT15-$V_{Ba}$ (dotted line) demonstrating very different macroscopic dielectric properties depending on the charge compensation scheme (a). The composition-dependent relative permittivity maximum ($T_c$ or $T_m$) for BNbT15-$V_{Ti}$ (solid line) and BNbT15-$V_{Ba}$ (dotted line) (b).

The above results not only prove that defects - especially vacancies - play a crucial role in the local structure, but also that they severely impact the resulting macroscopic dielectric properties. In fact, vacancies either at A or B site can have widely different properties from atomic to macroscopic scale (cf. Figure 5 and Figure 7). The rate at which $T_m$ is lowered as a function of $Nb^{5+}$ concentration in case of BNbT15-$V_{Ti}$ (cf. Figure 7-b) suggests that heterovalent substitution is an effective disruptor of long-range order not primarily because of ionic size difference, but rather because $V_{Ti}''''$-$Nb^{5+}$ defect complexes produce strong localized random fields. The difference in the diffusivity and sharp decrease in relative permittivity for BNbT15-$V_{Ti}$ can be related to the increase in the degree of heterogeneity at the B-site, which is the key crystallographic site that contributes to FE order (and its disruption) in BTO based systems.

## 2.5. Origin of Relaxor Behavior in Ba-Based Perovskites

To further explain the role of $V_{Ti}''''$-$Nb^{5+}$ defect clusters in BTO lattice, and to highlight the difference with relaxor behavior in homovalent substituted BTO systems, here represented by BZT, DFT calculations were carried out studying the effect of $Zr^{4+}$ or $Nb^{5+}$ addition (in the latter case with charge compensation by $V_{Ti}''''$) on the volume and total electric potential of the BTO lattice, as shown in **Figure 8**. 5x5x5 supercells were considered, in which 4 Ti atoms are replaced by 4 Zr atoms and 5 Ti atoms are replaced by 4 Nb atoms and one $V_{Ti}''''$, for BZT and BNbT, respectively (i.e. amounting to 3.2% substitution concentration for both BZT and BNbT). Note that Figure 8 is a stacked representation of 5 layers sliced from the 5x5x5 supercell for clarity.

To analyze the local strain caused by substituents, the local volume was calculated using the eight nearest neighboring Ba ions surrounding the B-site ions of each unit cell within the supercell. A comparison of these values to rhombohedral BTO yields a relative change in local volume (cf. Figure 8 (a) and (b) for BZT and BNbT, respectively). The change in total electric potential in BZT (c) and BNbT (d) was estimated for planes spanned by B-site ions by computing the difference to the pure BTO system. As can clearly be seen, in BZT there is a large strain difference in the vicinity of Zr atoms, impacting beyond the nearest-neighbor B-site, but the potential difference is confined to the Zr atom. In BNbT, on the other hand, the effect of the defect cluster is very strong both in terms of strain and electrical potential, and its effect is "felt" over a wide range of the supercell. This result again confirms the supposition that only in BNbT a charge-related mechanism is at the basis of the disruption of ferroelectric long-range order, and that these localized random fields can influence the direction of $Ti^{4+}$ and $Nb^{5+}$ cation displacements in several neighboring unit cells, inducing the strong lattice disorder we have detected by NMR and Raman, and the peculiar dielectric response, in BNbT systems. This disorder de facto disrupts the correlation of Ti-O-Ti chains, effectively hindering long-range ferroelectricity. On the other hand, in BZT only strain effects result from the difference in ionic radii and the extent of this effect is drastically smaller than the strain difference observed in BNbT due to the presence of $V_{Ti}''''$-$Nb^{5+}$ defect clusters. Hence, much higher substituent content is necessary in homovalent relaxors to effectively disrupt Ti-O-Ti chain correlation on the long-range, and thus to induce relaxor behavior. A direct analysis of the local dipoles gives additional evidence for the impact of Nb substitution compared to Zr substitution, as shown in Figure S9 in the supplemental material and the corresponding discussion. Figure S9 is inspired by the visualization of dipoles in defected supercells presented by Liu et al. [42], even though it is important to note that we deal only with a static result obtained at 0 K, while Liu et al. investigated the dynamical behavior of dipoles. Nevertheless, the analysis of local dipole changes also shows that Zr defects have a smaller impact on the neighboring dipole moments, and the impact is more isotropic, mostly changing the dipoles along the <100> directions. The Nb-$V_{Ti}$ clusters found in BNbT, in contrast, change the dipole moments of the neighboring cells much more strongly and anisotropically, where the strongest changes are found close to the titanium vacancy, which also determines the direction of the dipole changes.

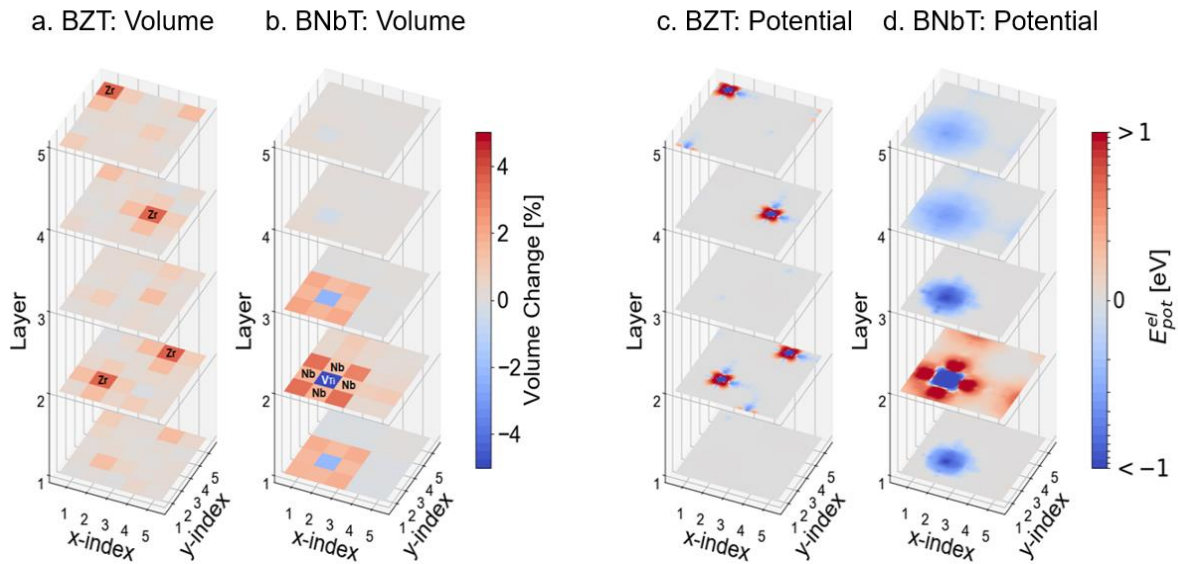

**Figure 8.** Local 5-atom unit cell volume and total electric potential change in BZT (a, c) and BNbT (b, d), respectively, as obtained from DFT calculations. Local volume change of each unit cell and the corresponding absolute change of total electric potential with respect to pure BTO is presented, showing the strong impact of $Nb^{5+}$ substitution on the local structure that extends to several unit cells compared to the weak and very localized influence of $Zr^{4+}$ substitution.

## 3. Conclusions

This study combines different experimental and theoretical approaches to show that the origin of relaxor behavior in lead-free BTO based systems is widely different for homovalent and heterovalent substituents. Charge compensation schemes, in fact, can result in different local structures and can impact differently the macroscopic material properties. The role of defects and defect distribution in disrupting the long-range ferroelectric order in heterovalent substitution of BTO is uncovered here through the analysis of $Nb^{5+}$-substituted BTO, demonstrating that the charge-mediated mechanism is more effective than a purely strain-mediated mechanism (as in homovalent $Zr^{4+}$-substituted BTO) in inducing ferroelectric lattice disruption, thereby resulting in the onset of relaxor behavior at lower substituent content. This understanding is important for the design of new perovskite materials for high energy density capacitors, since the recoverable energy density primarily relies on the reduction of the ferroelectric hysteresis. Judiciously modified BTO ceramics where a high lattice electrical polarizability is combined with the absence of long-range ferroelectric order, as the heterovalent system addressed in this work, may thus deserve attention as high-performance electrostatic energy storage materials.

## 4. Methods

### 4.1. Sample Preparation

$BaTiO_3$, $BaZr_xTi_{1-x}O_3$ and $BaNb_xTi_{1-5x/4}O_3$ ceramics were fabricated by the conventional solid-state mixed oxide route. Based on the required composition, stoichiometric amounts of pure $BaCO_3$ (electronic grade purity, Solvay Bario e Derivati, Italy; SSA = 3.3 $m^2$/g), $TiO_2$ (electronic grade purity, Toho, Japan; SSA = 6.1 $m^2$/g), $Nb_2O_5$ (99.9% purity, H.C. Starck,

SSA = 6.3 m$^2$/g, ceramic grade) and ZrO$_2$ (grade TZ0, Tosoh, Japan, SSA = 15.3 m$^2$/g) were weighted and wet mixed with zirconia grinding media using water with ammonium polyacrylate as additive for homogenous mixing. In the case of BNbT, where charge compensation is needed, Ti deficiency was taken into account to avail the formation of Ti vacancies ($V_{Ti}''''$, in Kroger-Vink notation). After calcination, the powders were compacted, pressed and sintered at different temperatures depending on the composition. Appropriate reoxidation annealing was performed for 48 h at 1000°C. More details on the procedure can be found in our previous work [8,36].

### 4.2. Weak Field Permittivity Measurements

The dielectric permittivity was measured in three different experimental setups based on the measuring frequency range. For 1 mHz to 1 MHz frequency range, the capacitance and loss tangent were measured with Solartron Modulab XM MTS system, which has a low current module. Additional experiments were carried out with HP4284A LCR meter in 20 Hz – 1 MHz frequency range. The dielectric permittivity in 1 MHz – 1 GHz frequency range was determined by measuring the complex reflection coefficient from the samples that are placed at the end of the coaxial line. An Agilent 8714ET vector network analyzer was used to measure the complex reflection coefficient. The complex dielectric permittivity was calculated according to the "multi-mode capacitor" model. A detailed description of this experimental setup can be found in the literature [43]. To determine permittivity in the 26 – 40 GHz frequency range, the sample was placed in a waveguide that supports only TE$_{10}$. A scalar network analyzer ELMIKA R2400 measured the scalar reflection and transmission coefficients simultaneously. The permittivity can be then determined by solving nonlinear equations using the modified Newton method. The details of this experimental setup can be found in the literature [44,45].

### 4.3. Temperature Dependent Raman Spectroscopy

Raman measurements were carried out in a LabRAM 300 spectrometer (Horiba Jobin Yvon, Villeneuve d'Ascq, France) using an excitation wavelength of λ = 532 nm in a backscattering geometry equipped with an edge filter (cut-off: 80 cm$^{-1}$), 1800 g/mm grating and CCD detector. The laser light was focused on the sample surface by means of a long working distance 100x objective (with NA 0.8, LMPlan FI, Olympus, Tokyo, Japan) and a spot size of 1 µm. The effective power at the sample surface was ~2 mW. Temperature-dependent Raman measurements were carried out in a Linkam temperature-controlled stage (THMS600, Linkam, Tadworth, UK) placed under the Raman microscope. The spectra were visualized using commercial software environment (Origin 2018b, OriginLab Corp., Northampton MA, USA) after correcting for the Bose-Einstein population factor.

**4.4. Nuclear Magnetic Resonance Spectroscopy**

$^{137}$Ba NMR spectra were recorded with a Bruker Avance III spectrometer at a frequency of 66.707 MHz. Ceramic samples were crushed to powders and annealed at 400 °C for two hours to relieve mechanical stress from grinding. A 3.2mm MAS probe was used in the variable-temperature, static, NMR experiments. A Hahn-echo pulse sequence was employed with 90° and 180° pulses of 2 µs and 4 µs, respectively, as well as an echo delay of 100 µs and a recycle delay of 100 ms. $^{137}$Ba NMR spectra were processed with a Gaussian window function no broader than 1/7 of the original linewidth and fitted with the Czjzek model implemented in DMfit [46]. From this procedure, the mean value for the distribution of quadrupolar coupling constants ($C_Q$) can be determined, and its error estimated by a statistical analysis of the fit and of the stability of the model through a Monte Carlo procedure with 100 attempts for each spectrum. Nominal temperatures are reported for the NMR spectra, which were recorded by a thermocouple positioned downstream from the sample.

**4.5. DFT Calculations**

The DFT calculations were performed using the VASP code [47–51]. We employed the PBEsol [52] functional to describe the exchange-correlation interactions. The energy cut-off of 520 eV was used for expanding the wave function in the basis of plane-waves. The sampling of the Brillouin zone was performed using the following Γ-centered uniform meshes: 8x8x8 for bulk, 3x3x3 for the 3x3x3 supercell, and 2x2x2 for both 4x4x4 and 5x5x5 supercells. As a preparatory step, we computed the converged structural parameters of the bulk rhombohedral BTO and obtained the lattice parameter of 4.009 Å and the angle α of ~89.86°, in good agreement to literature [53,54]. The supercells were then constructed from this optimized bulk r-BTO. After that, the supercells were fully relaxed, i.e., the ionic positions as well as the shape and the volume of the supercells were subject to optimization. The simulated Raman spectra were obtained using the spherical averaging method [20] employing 3x3x3 supercells.

## Supporting Information

Supporting Information is available from the Wiley Online Library or from the author.

## Acknowledgements

This project has received funding from the European Research Council (ERC) under the European Union's Horizon 2020 research and innovation program (grant agreement No 817190). V.V., M.P., J.S., and M.D., acknowledge funding also from the Austrian Science Fund (FWF): Projects P29563-N36 and I4581-N. S.S. has received funding from European Social Fund (project No. 09.3.3-LMT-K-712-19-0052) under grant agreement with the Research Council of Lithuania (LMTLT). P.B.G. acknowledges financial support by the Dutch Research Council (NWO) for the ECCM Tenure Track funding under project number ECCM.006, as well as the Deutsche Forschungsgemeinschaft (DFG) under contract Bu-911-28-1. M.P. and J.H. acknowledge the Czech Science Foundation (CSF project no. 20-20326L). Prof. Ronald J. Bakker (Montanuniversität Leoben, Chair of Resource Mineralogy) is gratefully acknowledged for providing access to Raman equipment. Prof. Gerd Buntkowsky is kindly acknowledged for providing access to the solid-state NMR facility at TU Darmstadt. Prof. Steven Tidrow (Kyocera Inamori Professor, New York State College of Ceramics, Alfred University, USA) is gratefully acknowledged for providing access to temperature dependent electrical characterization facilities at the Laboratory for Electroceramics (Alfred University) that were partly used by V.V. for this work.

## Conflict of Interest

The authors declare no conflict of interest.

## Author Contributions

V.V. performed Raman, dielectric permittivity measurements, treated the respective data, put together, plotted and interpreted the results from other techniques, and wrote the main part of the manuscript. M.N.P. conceived and performed all DFT calculations and computed simulated Raman spectra. F.M. evaluated DFT calculations for the study of local volume and total electric potential changes. J.S. actively contributed to concept, initiation and evaluation of ab-initio calculations, performed the eigenmode analysis and supported the overall interpretation of results. S.S. performed microwave dielectric spectroscopy experiments and analysis. V.K. and S.S. performed low temperature and low frequency dielectric spectroscopy experiments and analysis. J.B. organized the study of broadband dielectric spectroscopy. J.L. and P.B.G. performed the NMR measurements, along with the data analysis. P.B.G. contributed with the interpretation of NMR spectra, wrote initial passages related to the NMR data, and did the error analysis for spectral fitting. G.C. and M.T.B. prepared the samples and performed their basic characterization. V.B. organized sample preparation. M.P. supported the concept and development of DFT calculations. J.H. provided regular feedback and useful suggestions for modeling of Raman spectra. M.D. conceived, initiated the study and developed the concept, actively contributing to Raman data interpretation, overall results interpretation, and writing the discussion part of the manuscript. All authors corrected manuscript drafts.

Received: ((will be filled in by the editorial staff))
Revised: ((will be filled in by the editorial staff))
Published online: ((will be filled in by the editorial staff))

Vignaswaran Veerapandiyan, Maxim N. Popov, Florian Mayer, Jürgen Spitaler, Sarunas Svirskas, Vidmantas Kalendra, Jonas Lins, Giovanna Canu, Maria Teresa Buscaglia, Marek Pasciak, Juras Banys, Pedro B. Groszewicz, Vincenzo Buscaglia, Jiri Hlinka, Marco Deluca

**Origin of Relaxor Behavior in Barium Titanate Based Lead-Free Perovskites**

In $BaTiO_3$ solid solutions, relaxor behavior is caused by lattice disorder disrupting long-range ferroelectricity. In this work, we show that introducing donor substituents at the B-site of $BaTiO_3$ leads to the formation of clusters of charged defects (involving Ti vacancies) as a charge-compensation mechanism, which leads to an earlier onset of relaxor behavior compared to homovalent-substituted systems.

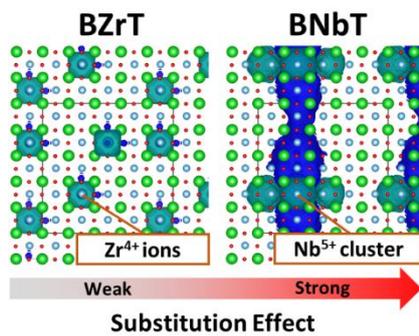

(Graphic for ToC)